\documentstyle[12pt]{article}

\newcommand{\fnorm}{\mbox{$\parallel\cdot\parallel_f$}}
\newcommand{\ffnorm}{\mbox{$\parallel\cdot\parallel_{\tilde f}$}}
\newcommand{\norm}{\mbox{$\parallel\cdot\parallel$}}
\newcommand{\hhbar}{\mbox{$\overline {\cal H}$}}
\newcommand{\cbar}{\mbox{$\overline {\cal C}$}}
\newcommand{\mcal}{\mbox{$\cal S$}}
\newcommand{\hcal}{\mbox{$\cal H$}}
\newcommand{\hcala} {\mbox{${\cal H}_1$}}
\newcommand{\hcalb} {\mbox{${\cal H}_2$}}
\newcommand{\ccal}{\mbox{$\cal C$}}

\newcommand{\lambdaz} {\mbox{$\stackrel{\circ}{\lambda}$}}
\newcommand{\muz} {\mbox{$\stackrel{\circ}{\mu}$}}

\newcommand{\qz} {\mbox{$\stackrel{\circ}{q}$}}
\newcommand{\Dz} {\mbox{$\stackrel{\circ}{D}$}}

\begin{document}
\title{Total Mass-Momentum of Arbitrary Initial-Data Sets in General
Relativity}
\author{Robert Geroch \and Shyan-Ming Perng \\
\em{ Enrico Fermi Institute, University of Chicago,} \\
\em {5640 S. Ellis Avenue, Chicago, Illinois  60637 } }
\maketitle

\begin{abstract}
For an asymptotically flat initial-data set in
general relativity, the total
mass-momentum may be interpreted as a Hermitian quadratic
form on the complex, two-dimensional vector space of
``asymptotic spinors''. We obtain a generalization to
an arbitrary initial-data set. The mass-momentum is retained as
a Hermitian quadratic form, but the space of ``asymptotic
spinors'' on which it is a function is modified. Indeed,
the dimension of this space may range from zero to infinity,
depending on the initial data. There is given a variety of
examples and general properties of this generalized mass-momentum.
\end{abstract}

\oddsidemargin=0in
\evensidemargin=0in
\topmargin=-.5in
\textwidth=6.7in
\textheight=9in
\hsize=6.27truein

\newpage

\section{Introduction}

There is a well known procedure$^1$  that assigns,
to any space-time that is asymptotically flat in a suitable sense,
a quantity representing the total mass-momentum of that
space-time, measured at spatial infinity.
To implement this procedure,
first draw in that space-time a spacelike slice $T$, and consider
the induced initial data --- consisting of the induced metric
$q_{ab}$ and the extrinsic curvature $p_{ab}$ --- on that slice.
Next, impose on these initial data asymptotic flatness:
that $q_{ab}$ approach a flat metric at infinity,
and $p_{ab}$ approach zero, at suitable rates.
Finally, define the components of the total mass-momentum
of the initial data as the values of certain asymptotic integrals,
whose integrands involve $q_{ab}$, its first derivative, and
$p_{ab}$. This procedure requires for its success a detailed
definition of asymptotic flatness, and the choice of definition is a
rather delicate business. On the one hand, the definition
must be weak enough that it permits $q_{ab}$ and $p_{ab}$ to
convey, in their asymptotic behavior, information about the
total mass-momentum. On the other hand, the definition
must be strong enough that it permits recognition
of ``asymptotic directions'' to serve as labels for the components
of the total mass-momentum.

There have been a number of attempts, over the past ten years
or so, to assign a suitable mass-momentum to a mere portion
of space-time. Consider, to be specific, initial data for
Einstein's equation given on a closed 3-ball $B$, with
2-sphere boundary $K$. Can there be defined something that
could reasonably be interpreted as ``the mass-momentum
within this region $B$''? Penrose$^2$
introduced a complex mass-momentum-angular-momentum given as
integrals over $K$ of spinor solutions of a certain
differential equation.
Dougan and Mason$^3$,
also using spinor integrals, introduced a real mass-momentum,
which turns out in addition to be timelike in an appropriate
sense.
Bartnik$^4$
used a different method to introduce a total mass (only)
associated with the region $B$: For each suitable extension of
the local initial data on $B$ to asymptotically flat, compute
in the usual way the total mass, and then minimize it over
extensions. There have also been proposed definitions involving
time evolution$^5$
of the data for short distance, and null evolution$^6$
all the way to null infinity.

Returning now to the asymptotically flat case, Witten$^7$
has obtained an elegant reformulation of this subject.
Introduce spinor fields on $(T,\, q_{ab})$. Consider now
a spinor field $\lambda^A$ that satisfies the
Witten equation --- a certain first order, neutrino-like
equation --- and that approaches a constant asymptotically.
Write down, for this $\lambda^A$,
a certain integral, (12), over $T$ with integrand quadratic
in $\lambda$ and its first derivative. It is known$^7$ that
this integral gives precisely the component of the
total mass-momentum vector in the asymptotic null
direction defined by the asymptotic behavior of $\lambda^A$.
Incidentally, the integrand for the mass integral is manifestly
non-negative, an observation that proves the positive-mass theorem.

This paper is based on two key observations regarding the
Witten reformulation:
first, that minimization of the mass integral yields automatically
the Witten equation; and, second, that finiteness of the mass integral
yields automatically that $\lambda^A$ approaches a constant
asymptotically. Thus, this single
integral is tied in to everything:
the equation the spinor field $\lambda^A$ is to
satisfy, the asymptotic conditions on that spinor field,
and the value of the mass-momentum component associated with
that spinor field. Indeed, the construction of the total mass-momentum
for an asymptotically flat initial-data set may be reformulated
as follows. First, introduce the space \mcal, essentially the
quotient of the space of
all spinor fields $\lambda^A$ on $(T,\ q_{ab})$ for which
the mass integral converges
by the subspace consisting of spinor
fields of compact support. In the present instance, this
\mcal\ will be a two-dimensional
vector space of ``asymptotic spinors'',
and so will provide a space of ``directions along
which to evaluate the components of the mass-momentum''.
Next, introduce on this space \mcal\ the function, $M$, whose
value is given by the minimum of the mass integral.
In the present instance, this $M$ will provide the appropriate
mass-momentum components.

But note that the formulation of the previous paragraph nowhere
uses asymptotic flatness. Thus quite generally --- for any
initial-data set whatever --- one can introduce
the quotient, \mcal, of spinor fields with finite mass integral
by those of compact support, and then introduce the function
$M$ on \mcal\ by minimizing the mass integral. For asymptotically flat
initial data, this construction yields  the usual mass-momentum
components.
What happens for more exotic initial data ---
say, with $T$ compact, or consisting of a small patch
from a large initial-data set?
In general, the space \mcal\ becomes modified in some way --- it is
no longer a simple two-dimensional vector space.
Indeed, its dimension can range from zero to infinity.
Thus, we retain the
mass-momentum as a function on possible component-directions, but
the space of such directions becomes more complicated.
This, we suggest, is the natural notion of ``total energy-momentum''
for a general initial-data set.

This paper is organized as follows.
Sect. 2 contains the basic definitions. We first introduce
the space \mcal\ representing the ``asymptotic spinors''.
It turns out that there are actually two natural mass
functions on this space \mcal\ --- what we call the
norm mass function $M_N$ and the asymptotic mass function $M_A$.
We show that, in the case of asymptotically flat initial data, the
space \mcal\ reduces to a two-dimensional vector space,
while the two mass functions coincide and yield the appropriate
components of the total mass-momentum. Sect. 3 contains various
examples and properties of the space \mcal\ and the mass functions. The
space \mcal\ collapses to a single point --- and the mass
functions then necessarily to zero --- when there is
either ``too much matter'' or ``too little room asymptotically''.
By contrast, there is a large class of examples in which \mcal\
is infinite-dimensional, with rather complicated
mass functions.
For virtually all complete initial data, the
two mass functions are equal, while in the incomplete
case the two can differ.
Indeed, the asymptotic mass function can become negative in
certain cases, while the norm mass function never can.
The asymptotic mass function ---
but in general not the norm mass function --- depends only on the
``asymptotic behavior'' of the initial data. Finally, we
show that, for initial data with several ``asymptotic regions'',
there is a decomposition of the space \mcal\ and of
the mass functions into pieces associated with
the individual asymptotic regions.

\section{Basic definitions}

Fix an {\em initial-data set}. That is, fix a connected, 3-dimensional
manifold $T$, a smooth positive-definite metric $q_{ab}$ on $T$ and a smooth
symmetric tensor field $p_{ab}$ on $T$. Given such an initial-data set, we
set
\begin{eqnarray}
{\rho}&=&{1\over 2}\ [{\cal R} - p_{ab} p^{ab} +(p^m_{\ m})^2],\\
{\rho}_a&=&D^b(p_{ab}-p^m_{\ m}q_{ab}),
\end{eqnarray}
where ${\cal R}$ is the scalar curvature, and $D_a$ the derivative operator,
with respect to the metric $q_{ab}$ on $T$.
These will be recognized$^8$ as
the mass and momentum density, respectively, of the matter source.
The {\it energy condition} on this initial-data set is the
condition that the mass-momentum vector be future-directed non-spacelike:
\begin{equation}
\rho \geq ({\rho}^a{\rho}_a)^{1/2}.
\end{equation}
We shall here deal only with initial-data sets satisfying the energy
condition.

It is necessary for what will follow to introduce the notion of
spinor fields on such an initial-data set. To this end, fix a complex,
two-dimensional vector space $V$. By a spinor, we mean any element of a
tensor product involving $V$,
its complex-conjugate space $\overline V$,
and their respective dual spaces, $V^*$ and ${\overline V}^*$.
We designate spinors by upper-case Latin
indices --- unprimed superscripts for $V$,
primed superscripts for
$\overline V$,
and corresponding subscripts for their
corresponding duals.
Thus, ${\alpha}_{\quad ART'}^{BD'}$
( an element of
$V\otimes {\overline V} \otimes V^*
\otimes V^* \otimes {\overline V}^*$)
is a typical spinor.
By construction, we have on spinors
the operations of outer product, contraction,
complex conjugation (denoted by a bar),
and, for spinors with identical
index structure, addition.
For example, if ${\alpha}_A$,\ ${\beta}^B_{\ \,C'}$
and ${\gamma}_D$ are spinors, then so is
${\alpha}_A{\beta}^A_{\ \,C'}+\overline {\gamma}_{C'}$.
Now fix any antisymmetric spinor $\epsilon_{AB}$, and any real,
positive-definite spinor $t_{AA'}$
(i.e. $t_{AA'}\psi^A \overline {\psi}^{A'}$\ is
real and positive for any nonzero $\psi^A$),
with these normalized with respect to
each other by
\begin{equation}
t_{A[A'}t_{B']B}={1\over 2}\epsilon_{AB} {\overline \epsilon}_{A'B'}.
\end{equation}
These two fixed spinors are incorporated into
the notation in the following way.
Use $t_{AA'}$ and its inverse $t^{AA'}$(whose
existence is guaranteed by positive-definiteness)
to eliminate primed spinor indices in favor of unprimed ones.
Thus, we need only work throughout with
unprimed spinor indices.
Then use $\epsilon_{AB}$, and its inverse
$\epsilon^{AB}$(whose existence is guaranteed by
Eqn. (4)) to lower and raise these unprimed
spinor indices in the usual way
(i.e., $\kappa_A=\kappa^B \epsilon_{BA},\
\kappa^A=\epsilon^{AB}\kappa_B$ ).
The operation of complex conjugation on general
spinors then translates to an adjoint operation
on these unprimed spinors:
\begin{equation}
{\alpha^{A...C}_{\qquad B...D}}^{\dagger}
=(-1)^s t^A_{\ \ A'}...t^C_{\ \ C'}
t_B^{\ \ B'}...t_D^{\ \ D'}
{\overline \alpha}^{A'...C'}_{\qquad B'...D'},
\end{equation}
where $s$ is the number of superscripts of $\alpha$.
Thus, the adjoint of a scalar
is its complex conjugate, while,
$\epsilon_{AB},\ \epsilon^{AB}$, and the
unit spinor $\delta^A_{\ \,B}$ are self-adjoint.
This adjoint operation commutes
with outer product and contraction, and so with the raising
and lowering of indices.
Further, we have $\dagger\dagger =(-1)^r$,
where $r$ is the number of indices of the spinor to which
this equation is applied.
It follows from positive-definiteness
of $t_{AA'}$ that,
for every spinor $\alpha^{A...C} $,
$\ |\alpha|^2=(\alpha^{A...C})^{\dagger}\alpha_{A...C}\ge 0$,
with equality if and only if $\alpha =0$.
But note, e.g., that
$\alpha^{\dagger}_{\ A}\alpha^A \le 0$.
The set of symmetric, self-adjoint spinors
$\lambda^{AB}$  forms a real
3-dimensional vector space,
on which $\lambda^{AB}\lambda_{AB}$ is a positive-definite norm.
Now identify this vector space
with the tangent space at each point of $T$,
in such a way that this positive-definite
norm corresponds to the norm on the tangent
space arising from $q_{ab}$.
There results
the notion of {\it spinor fields} on $T$. Thus, each tensor field
on the manifold $T$ gives rise to a spinor field with twice as many
(spinor) indices.
Other examples of spinor fields include
$\epsilon_{AB}$,\
$\epsilon^{AB}$\ and
$\delta^A_{\ \,B}$.
The operations of outer product, contraction,
addition, and taking of the adjoint extend
immediately from spinors to spinor fields.
The tensor field
$q_{ab}$ on $T$ gives rise to the spinor field
$q_{ABCD}=q_{(CD)(AB)}=-\epsilon_{A(C}\epsilon_{D)B}$
on $T$. Finally,
the derivative operator $D_a$ on tensor fields on $T$ gives rise to a unique
corresponding derivative operator
$D_{AB}$ on spinor fields on $T$.
This operator is
symmetric in its indices,
satisfies the Leibnitz rule under outer
product,
commutes with
addition, contraction, and
the adjoint operation, and satisfies
$D_{AB}\epsilon_{CD}=0$.

So far, we have used only the metric $q_{ab}$ of $T$, and not the
symmetric tensor $p_{ab}$. We incorporate the latter by introducing a new
operator, ${\cal D}_{AB}$, on spinor fields, with action
\begin{eqnarray}
{\cal D}_{AB}\lambda_{C\ldots}^{\quad D\ldots}
=D_{AB}\lambda_{C\ldots}^{\quad D\ldots}
&+&{i\over {\sqrt 2}} {p_{ABC}}^M \lambda_{M\ldots}
^{\quad D\ldots}+\cdots \nonumber \\
&-&{i\over {\sqrt 2}}{p_{ABM}}^{D}\lambda_{C\ldots}
^{\quad M\ldots}-\cdots .
\end{eqnarray}
Here, $p_{ABCD}=p_{(CD)(AB)}=p^\dagger_{ABCD}$
is the spinor representation of $p_{ab}$.
For example, we have
\begin{eqnarray}
{\cal D}_{AB}\lambda^B&=&D_{AB}\lambda^B
+{i\over {2\sqrt 2}}p\lambda_A \\
{\cal D}_{AB}w^{AB}&=&D_{AB}w^{AB}
\end{eqnarray}
where we have set $p =p^{AB}_{\quad AB}=p^m_{\ \ m}$.
This operator ${\cal D}_{AB}$ shares with $D_{AB}$
the Leibnitz rule, annihilation of $\epsilon_{AB}$,
and commutation with addition, contraction and
raising and lowering of indices.
But ${\cal D}_{AB}$, in contrast to $D_{AB}$,
fails to be torsion-free
and also fails to commute with the adjoint operation.
Indeed, we have
\begin{equation}
({\cal D}_{AB}\lambda_{C}^{\ \ D})^\dagger
={\cal D}_{AB}\lambda^{\dagger \ D}_{C}
-{2i\over {\sqrt 2}} {p_{ABC}}^{M}\lambda ^{\dagger \  D}_{M}
+{2i\over {\sqrt 2}}{p_{ABM}}^{D}\lambda^{\dagger \  M}_{C}.
\end{equation}
It is convenient to introduce the
adjoint of ${\cal D}_{AB}$, defined as follows:
\begin{equation}
({\cal D}_{AB}\lambda^{\ \quad D\ldots}_{C\ldots})^\dagger =
{\cal D}^\dagger_{AB} \lambda_{C\ldots}^{\dagger\quad D\ldots}\ .
\end{equation}
What makes the operator ${\cal D}_{AB}$
so useful is the Witten-Sen identity$^9$,
which we shall use repeatedly:
For any spinor fields $\sigma^A,\ \lambda^A$ on $T$,
we have
\begin{eqnarray}
&&({\cal D}^{AB}\sigma^C)^\dagger({\cal D}_{AB}\lambda_C)+{1\over 2}
(\rho\sigma^{\dagger A}\lambda_A
+i\sqrt 2\rho_{AB}\sigma^{\dagger A}\lambda^B) \nonumber \\
&=&{\cal D}^{AB}(\sigma^{\dagger C}{\cal D}_{AB}\lambda_C)
 -2\sigma^{\dagger C}{\cal D}^{\dagger A}_{\quad C}{\cal D}_{AB}\lambda^B
\nonumber \\
&=&{\cal D}^{AB}(\sigma^{\dagger C} {\cal D}_{AB}\lambda_C
+2\sigma^\dagger_{\ B} {\cal D}_{AC}\lambda^C)
+2({\cal D}^A_{\ \ B}\sigma^B)^\dagger
({\cal D}_{AC}\lambda^C).
\end{eqnarray}
To prove Eqn. (11), expand
the right hand sides
using the Leibnitz rule, eliminate
${\cal D}_{AB}$ in favor of $D_{AB}$
using Eqn. (6), eliminate all second derivatives
using
$D_{M(A}{D_{B)}}^{M}\lambda^B
=\displaystyle{{{\cal R}\over 8}}\lambda_A$,
and finally eliminate
the $p_{ab}$'s using Eqns. (1) -- (2).

Now fix any initial-data set satisfying the energy condition (3).
Denote by ${\cal H}$ the collection of all smooth spinor fields $\lambda_A$ on
$T$ for which
the right side of
\begin{equation}
\parallel\lambda\parallel^2
\equiv \int_T \Bigl \{ ({\cal D}^{AB}\lambda^C)^\dagger
({\cal D}_{AB}\lambda_C)+{1\over 2}
(\rho\lambda^{\dagger A}\lambda_A
+i\sqrt 2\rho_{AB}\lambda^{\dagger A}\lambda^B) \Bigr \}
\end{equation}
converges. Note that this right side is nonnegative. Indeed,
the first term in the integrand on the right is manifestly nonnegative,
while the second
term is nonnegative because of the energy condition and the fact that the
norm of the (real) vector $i{\lambda^\dagger}_{(A} \lambda_{B)}$
is $\displaystyle{{1\over \sqrt 2}}\lambda^{\dagger A}\lambda_A$.
Thus, Eqn. (12) defines a
quadratic, positive semi-definite, norm on ${\cal H}$.
It follows from this that ${\cal H}$ is a (complex)
vector space.

We next construct a certain completion, \hhbar,
of ${\cal H}$. This is to be done
essentially via the norm (12) --- but we must exercise some care
to accommodate the fact that
this norm need not be strictly positive-definite.
Suppose for a moment that
there were some point {\it x} of $T$ at which
$\rho$ strictly exceeded $(\rho^a\rho_a)^{1/2}$.
Then, the norm (12) would already be
strictly positive-definite. Indeed,
$\parallel\!\lambda\!\parallel^2=0$ would imply,
by the right side of (12), the vanishing of
of $\lambda_A$ at $x$ and of
${\cal D}_{AB}\lambda_C$ everywhere.
But these two together
imply the vanishing of $\lambda_A$ everywhere.
Thus, in this case --- when there
is some point $x\in T$ at which $\rho > (\rho^a\rho_a)^{1/2}$
--- the norm (12) is already strictly positive-definite,
and so we may simply take for \hhbar\
the completion of ${\cal H}$ in this norm.
But what if there exists
no such point $x$ ?
In this case, we introduce a new norm,
$\parallel\cdot\parallel_f$,
obtained by adding to $\rho$ in (12)
any nonnegative function, $2f$,
somewhere strictly positive, of compact support:
\begin{equation}
\parallel\!\lambda\!\parallel^2_f
= \parallel\!\lambda\!\parallel^2
+\int_T{f|\lambda|^2}.
\end{equation}
This norm is automatically positive-definite,
and so we take for \hhbar\ the completion of \hcal\ in it.
The result is independent of the choice of the function $f$:

{\bf Theorem 1:}  Let $f$, $\tilde f$\ be two
functions on $T$,
each of which is nonnegative, somewhere strictly positive,
and of compact support. Then each of the two
norms \fnorm\ and \ffnorm\
bounds some multiple of the other.

{\it Proof\/}:
Fix $\lambda^A \in {\cal H}$. Let $w^a$ be any smooth vector
field on $T$ of compact support, denote by $\zeta_t\ (t\in R)$
the corresponding one-parameter family of diffeomorphism on $T$,
and set $\alpha (t) =\int_T{ (\zeta_tf) |\lambda|^2}$.
Then we have
\begin{eqnarray}
{d\alpha \over dt}
&=&\int{w^a \bigl [ D_a (\zeta_tf) \bigr ]|\lambda|^2}\nonumber\\
&=&\int \Bigl [-(D_aw^a)(\zeta_tf)|\lambda|^2
 -(\zeta_tf) w^a D_a |\lambda|^2 \Bigr ]\nonumber\\
&=&\int
\Bigl [ -(D_aw^a)(\zeta_tf)|\lambda|^2 -i\sqrt 2(\zeta_tf)
w^{AB}p_{ABCD}\lambda^C\lambda^{\dagger D} -(\zeta_tf)w^{AB}
\nonumber \\
&& \qquad \times \bigl (\lambda^{\dagger C}{\cal D}_{AB}\lambda_C
 -\lambda^C ({\cal D}_{AB}\lambda_C)^\dagger \bigr)\Bigr ]
\nonumber \\
&\le &\int \Bigl [(-D_aw^a+|p_{ab}w^b|)(\zeta_tf)|\lambda|^2 \Bigr ]
+2 \biggl [ \int {|w|^2(\zeta_tf)|\lambda|^2}
 \int{|{\cal D}_{AB}\lambda_C|^2} \biggr ]^{1/2}\nonumber \\
&\le & b\alpha +c\alpha^{1/2} \biggl [ \int{|{\cal D}_{AB}\lambda_C|^2}
\biggr ]^{1/2},
\end{eqnarray}
where $b$ and $c$ are positive numbers independent of
$\lambda^A$.
In the last step, we used the
Schwarz inequality. Solving this differential
inequality, we learn that $\alpha(t)$
is bounded by a linear combination, with
coefficients independent of $\lambda^A$, of
$\alpha(0)$ and $\int{|{\cal D}_{AB}\lambda_C|^2}$.
Hence, some multiple of
$\parallel\!\lambda\!\parallel_f$
bounds
$\parallel\!\lambda\!\parallel_{\zeta_tf}$.
The result now follows from the fact that
$\tilde f$ is bounded by a finite linear combination
of functions of the form
$\zeta_tf$.\ \ /

It follows from Theorem 1 that
any sequence Cauchy
in the norm $\parallel\cdot\parallel_f$
is also Cauchy in the norm $\parallel\cdot\parallel_{\tilde f}$.
, and therefore that the completion, \hhbar,\
of \hcal\ is indeed independent of
the function $f$ used to take that completion.
This $\overline {\cal H}$ is,
by construction, a complete
topological vector space with continuous,
quadratic, positive semi-definite norm \norm.

An element of $\overline {\cal H}$
is represented, via the construction above,
by a sequence of spinor fields $\lbrace\lambda_A^i\rbrace$
on $T$, Cauchy in the norm $\parallel\cdot\parallel_f$.
But there exists a more explicit representation.
To obtain it, let $\lbrace\lambda_A^i\rbrace$ be
any Cauchy sequence in \hcal\ and $U$ any
open subset of $T$ with compact closure.
Then the sequences
$\lbrace\lambda_A^i\rbrace$
and $\lbrace{\cal D}_{AB}\lambda_C^i\rbrace$
are both Cauchy in $L^2(U)$, by Theorem 1
and Eqn. (12) respectively, and therefore
converge in $L^2(U)$ to some
spinor fields $\kappa_A$ and
$\omega_{ABC}$ respectively. Furthermore,
this $\omega_{ABC}$ is actually the {\it weak derivative}
of  $\kappa_A$, i.e., we have,
for every smooth $\tau^{ABC}$ of compact support in $U$,
\begin{equation}
-\int_U{({\cal D}_{AB}\tau^{ABC})\kappa_C}
=\int_U{\tau^{ABC}\omega_{ABC}}.
\end{equation}
To see this, note that
\begin{eqnarray}
&&\biggl |\int_U \Bigl [\tau^{ABC}\omega_{ABC}
+({\cal D}_{AB}\tau^{ABC})\kappa_C \Bigr ]
\biggr|\nonumber \\
&=& \biggl |
\int_U \Bigl [\tau^{ABC}(\omega_{ABC}-{\cal D}_{AB}\lambda^i_C)
+({\cal D}_{AB}\tau^{ABC})(\kappa_C-\lambda^i_C)\Bigr ]
\biggr | \nonumber \\
&\le & \biggl [ \int_U{|\tau|^2}\int_U{|\omega_{ABC}
-{\cal D}_{AB}\lambda^i_C|^2}
\biggr ]^{1/2}
+\biggl [\int_U{|{\cal D}_{AB}\tau^{ABC}|^2}
\int_U{|\kappa_C-\lambda^i_C|^2} \biggr ]^{1/2},
\end{eqnarray}
while the right side approaches zero as
$i$ approaches infinity.
Since the subset $U$, open with compact closure,
is otherwise arbitrary, we conclude:
Each element of $\overline {\cal H}$
can be represented uniquely by a spinor field on $T$,
locally square integrable
with locally square-integrable
weak first derivative,
for which the integral (12) converges.
Then Eqn. (12) gives the
continuous, positive semi-definite norm
on this  $\overline {\cal H}$.
In the ``generic case''
--- when there is some point of $T$ at which
$\rho > (\rho^a\rho_a)^{1/2}$ ---
this $\overline {\cal H}$
is actually a Hilbert space under this norm.

Next, denote by $\cal C$ the collection of all smooth spinor fields
$\lambda_A$ on $T$ of compact support. Since every such spinor field is
automatically in $\cal H$  , we have that $\cal C$ is a complex vector subspace
of
$\cal H$, and so also of its completion $\overline {\cal H}$. Denote by
$\overline {\cal C}$
the closure of $\cal C$ in $\overline {\cal H}$,
so $\overline {\cal C}$ is a closed
subspace of  $\overline {\cal H}$.
Thus, an element of $\overline {\cal C}$
is also represented by
a Cauchy sequence of smooth spinor fields $\lambda_A$
of compact support.
But note that the corresponding
locally square-integrable limiting
spinor field, obtained as above, need not have compact support.
Finally, denote
by \mcal\ the quotient $\overline{\cal H}/\,\overline{\cal C}$,
so \mcal\ is itself a complete topological vector space.
Thus, an element of \mcal\ is
represented by a Cauchy sequence of smooth spinor fields
$\lambda_A$ on $T$, where two such sequences define the same element of \mcal\
provided
their difference converges to some element of $\overline
{\cal C}$.
Alternatively, an element of \mcal\ may be represented by
a spinor field in \hhbar, where two spinor fields define the same element
of \mcal\ provided their difference is in \cbar.

We now introduce two functions, $M_N$ and $M_A$, on \mcal\
as follows. Fix $\alpha \in$ \mcal,
and consider
\begin{eqnarray}
M_N(\alpha)&=&\inf \parallel\!\lambda\!\parallel^2, \\
M_A(\alpha)&=& \parallel\!\lambda\!\parallel^2 -2\int_T{
({\cal D}^A_{\ B}\lambda^B)^\dagger ({\cal D}_{AC}\lambda^C)}.
\end{eqnarray}
In the first, the infimum on the right is over all
$\lambda$ in the equivalent class $\alpha$,
and so this right side indeed yields a function, $M_N$, on \mcal.
For the second, first note that the right side is a
continuous function on \hcal\ ( by
$|{\cal D}_{AB}\lambda^B|^2 \le
{3\over 2}
|{\cal D}_{AB}\lambda_C|^2$\ ) that
vanishes on \ccal\ ( by Eqn. (11)).
Hence, that right side extends continuously
to a function on \hhbar\ that vanishes on \cbar,
thus yielding a function, $M_A$, on \mcal.
Note that
we may evaluate that right side of Eqn. (18) for any $\lambda$
in the equivalence class $\alpha$.
Both$^{10}$ of the functions are continuous
and quadratic$^{11}$.
For reasons that will emerge shortly,
we call $M_N$\
the {\it norm mass function}, and
$M_A$\ the {\it asymptotic mass function}.

The following example will motivate and illustrate
these definitions.
Let $T = R^3$, let $q_{ab}$ be the usual Euclidean metric on $R^3$, and let
$p_{ab}=0$.
This is initial data for Minkowski spacetime.
 From Eqns. (1) -- (2), these data have $\rho=0$ and $\rho_a=0$,
and so satisfy the energy condition.
We shall show that,
for this initial-data set,
\mcal\
is a 2-dimensional complex
vector space, which may be identified
with the space of constant spinor fields
on $T$, while
both mass functions vanish.

Which smooth spinor fields $\lambda_A$
on $T$ have finite norm (12), i.e., which are in
$\cal H$? We first
show that every such $\lambda_A$   must,
in a suitable sense, approach a constant
asymptotically.

{\bf Theorem 2:} Let $(T, q_{ab}, p_{ab})$\ be the
above initial data
for Minkowski space-time, and let $\lambda_A \in \cal H$.
Then there exists a constant spinor field
\lambdaz$_A$
on $T$ such that
\begin{equation}
\int_T{ \Bigl |{1\over r}(\lambda_A-{ \lambdaz_A }) \Bigr |^2}
\le {9\over 2}\parallel\!\lambda\!\parallel^2,
\end{equation}
where $r$ denotes distance from some fixed origin on $T$.

{\it Proof\/}: In the norm (12) in this case, only
the first term on the right survives.
Taking one component at a time, it suffices
to prove the result for a smooth scalar field $\lambda$\ ,
with
\begin{equation}
\parallel\!\lambda\!\parallel^2=\int_T{|D\lambda|^2}
\end{equation}
finite. For each $0\le r <\infty$\ , set
\begin{equation}
g(r)=\int_{S_r}{\lambda d\Omega},
\end{equation}
where $S_r$ denotes the sphere of radius $r$ centered at the fixed
origin, and $d\Omega$ its unit surface-element. We have
\begin{eqnarray}
r^2 \biggl ( {{dg}\over{dr}} \biggr )^2
&=&\biggl [ r \int_{S_r}{(D_a\lambda)
(D^ar)d\Omega} \biggr ]^2 \nonumber \\
&\le & \int_{S_r}{ r^2 (D^a\lambda D_a\lambda) d\Omega}
\int_{S_r}{(D^a r D_a r) d \Omega} \nonumber \\
&=&\int_{S_r}{|D\lambda|^2r^2d\Omega\cdot{4\pi}}.
\end{eqnarray}
Integrating over $r$, we obtain
\begin{equation}
\int_0^\infty{r^2\Bigl ({{dg}\over{dr}}\Bigr )^2dr}
\le 4\pi \parallel\!\lambda\!\parallel^2.
\end{equation}
It follows from (23), since
$\parallel\!\lambda\!\parallel^2$ is finite,
that $g(r)$
has a limit as $r$ approaches infinity. Subtract a constant,
\lambdaz ,
from $\lambda$\ so that this limit
becomes zero. Now expand $\lambda$ on $S_r$ in
spherical harmonics to obtain
\begin{equation}
\int_{S_r}{\lambda^2 d\Omega}\le
{{r^2}\over 2}\int_{S_r}{|D\lambda|^2d\Omega}
+{1 \over {4\pi}} \biggl [ \int_{S_r}{\lambda d\Omega} \biggr ]^2.
\end{equation}
Integrating this inequality over $r$, the first term on
the right is bounded by $\displaystyle{1\over 2}
\parallel\!\lambda\!\parallel^2$, and the second
by $4\parallel\!\lambda\!\parallel^2$,
where we have used for the latter (23) and the
following fact:
If $g(r)$ approaches 0 as $r$ approaches infinity, then
\begin{equation}
\int_0^\infty {g^2 dr}\le 4\int_0^\infty
{r^2 \bigl ( {{dg}\over{dr}} \bigr )^2dr}.\ \ \ /
\end{equation}

The constant spinor field
\lambdaz$_A$
whose existence is guaranteed by the theorem is of course
unique given $\lambda_A$.
The theorem says, roughly speaking,
that $\lambda_A$     approaches
\lambdaz$_A$
`` faster than $r^{-1/2}$''.
Thus, Theorem 2 gives the asymptotic behavior
of the spinor fields with
finite norm (i.e., those in $\cal H$). Which of these are limits of spinor
fields of compact support, i.e., which are in \cbar? The answer is
provided by the following.

{\bf Theorem 3:}  Let, as in Theorem 2, $\lambda_A \in \cal H$.
Then this $\lambda_A$ is in \cbar
\ if and only if the constant field
\lambdaz$_A$
of that theorem vanishes.

{\it Proof\/}: For the ``if'' part, let this
$\lambda_A$\ have
$\lambdaz_A=0$.
Fix any number
$r_0>0$, and any smooth nonnegative function
$h(r)$\ with $h(r)=1$\ for $r<r_0$, $h(r)=0$\ for $r>2r_0$,
and $|dh/ dr|\le 2/r$\ for all r.
Then $h(r)\lambda_A$\ has compact support, while
\begin{eqnarray}
\parallel\!\lambda_A-h\lambda_A\!\parallel^2
&=&\int_T{ \Bigl |D[(1-h)\lambda_A] \Bigr |^2}\nonumber \\
&\le & 2\int_T \Bigl [(1-h)^2|D\lambda_A|^2 +|\lambda_A|^2|Dh|^2 \Bigr ]
\nonumber \\
&\le & 2\int_{r\ge r_0} { \Bigl [ |D\lambda_A|^2
+4{{|\lambda_A|^2}\over {r^2}} \Bigr ]}
\end{eqnarray}

The last integral above converges, by $\lambda_A\in {\cal H}$ and
Theorem 2 with $\lambdaz_A=0$, and the integrand is independent of $r_0$,
so that integral approaches zero as $r_0$ approaches infinity.
Repeating this argument for a succession of values of $r_0$,
approaching infinity, we obtain a sequence of spinor fields of compact
support, the corresponding $h\lambda_A$'s, which, by (26),
converge to $\lambda_A$. For the ``only if'' part,
fix any $r_0>0$ and any smooth vector field
$w^a$\ on $T$ equal to $-\displaystyle{{1\over {4\pi}}}D^a(1/r)$
for $r>r_0$. Then, for any $\mu_A\in {\cal H}$
we have, again suppressing the spinor index,
\begin{eqnarray}
\muz &=&\int_T{D_a(\mu w^a)} =\int_T{w^aD_a\mu + \mu D_aw^a}\nonumber \\
&\le & \biggl [ \int_T{w^aw_a}\int_T{D^a\mu D_a\mu} \biggr ]^{1/2}
+\biggl [ \int_{r\le r_0}{\mu^2}\int_{r\le r_0}{(D_aw^a)^2} \biggr ]^{1/2}
\end{eqnarray}
But this formula shows that
$\lambdaz_A$,
the
asymptotic value of $\lambda_A$,
is
continuous in the topology of $\cal H$. The
result follows. /

Thus, two elements of $\cal H$ differ by an element of \cbar \ when and
only when they approach the same constant spinor field asymptotically.
It follows that, in this example,
\mcal(=\hhbar$/\,$\cbar)
is a (complex) 2-dimensional vector space,
which  may be identified
with the space of constant spinor fields on $T$.
The operations of taking the adjoint and taking the $\epsilon$-inner
product on constant spinor fields on $T$ extend, via this
identification, to corresponding operations on \mcal. Thus \mcal\
has all the structure of a spinor space. Both of the
mass functions on \mcal\ are zero,
since the right sides of (17) and (18)
vanish for constant $\lambda^A$.
This, then, is \mcal\ and the mass functions for these data for
Minkowski space-time.

We turn now from flat initial data to asymptotically flat.
We shall see that in this case the present framework
yields the physically correct answer, an observation that
serves as motivation for this framework. We first show
that, for an initial-data set asymptotically flat in
a suitable sense, the space \mcal\  has a structure
identical to that for Minkowski initial data.

{\bf Theorem 4:}$^{12}$
Let $(T,\, q_{ab},\, p_{ab})$ be an initial-data set satisfying the
energy condition, and having $T=R^3$. Let $\qz_{ab}$ be a
Euclidean metric on $T$ such that

$i)$ $q_{ab}-\qz_{ab}$ approaches$^{13}$
zero asymptotically;

$ii)$ the fields $p_{ab}$ and $\Dz_a q_{bc}$ are square-integrable,
and the source $\rho$ integrable, over $T$; and

$iii)$ both $rp_{ab}$ and $r\Dz_a q_{bc}$ approach$^{13}$ zero
asymptotically,

\noindent where $\Dz_a$ denotes the $\qz$-derivative
operator, and $r$ $q$-distance from some origin. Then

$i)$ the space \mcal\ is 2-dimensional.

$ii)$ each element, $\alpha$, of \mcal\ has representative,
$\lambda^A$, such that $\lambda^{\dagger A}$ is also in \hcal; and

$iii)$ any two elements, $\alpha$ and $\beta$, of \mcal\
have representatives, $\lambda^A$ and $\mu^A$, such that
the function $\lambda_A\mu^A$ approaches$^{13}$ a constant asymptotically.

{\it Proof\/}: Let \lambdaz$^A$ be any constant spinor
field on $(T,\, \qz_{ab})$. Denote by $\lambda^A$ that spinor
field (unique up to sign) on $(T,\, q_{ab})$ such that,
at each point of $T$, the real part of the complex vector
$\lambda^A\lambda^B$, as well as the 2-plane spanned by its
real and imaginary parts, are identical with the corresponding
vector and plane for \lambdaz$^A$\lambdaz$^B$. It is immediate
from hypotheses $i)$ and $ii)$ that the $\lambda^A$ so constructed
is in \hcal\ (and so is the representative  of some element
of \mcal), and that the representatives so obtained themselves
already satisfy conclusions $ii)$ and $iii)$ of the theorem.
Thus, there remains only to show that every element of \mcal\
is obtained via this construction, and that the zero element
of \mcal\ is obtained only via \lambdaz$^A=0$.
These, in turn, are proven along the lines of
Theorems 2 and 3, so modified to retain, and
then bound via hypotheses $ii)$ and $iii)$, the additional
terms involving $p_{ab}$ and $\Dz_a q_{bc}$.

Fix $\mu^A\in$\hcal, denote by ``$\lambda$'' its components with
respect to a basis constructed as in the paragraph above,
and again define $g(r)$ by Eqn. (21). Then Eqn. (22)
is replaced by
\begin{equation}
r^2 \Bigl( {{dg}\over{dr}} \Bigr )^2 \le
8 \pi r^2 \int_{S_r} |{\cal D}_{AB} \mu_C|^2 d\Omega
+ r^2 V(r)\int_{S_r} \lambda^2 d\Omega ,
\end{equation}
where ``$V(r)$'' denotes the integral over $S_r$ of a certain
expression quadratic in $p_{ab}$ and $\Dz_a q_{bc}$.
So $r^2 V(r)$ is bounded, by hypothesis $iii)$, and
$r$-integrable to $r=\infty$, by hypothesis $ii)$.
Eqn. (24) is replaced by
\begin{equation}
\int_{S_r} \lambda^2 d\Omega \le 2r^2 \int_{S_r}
|{\cal D}_{AB}\mu_C|^2 d\Omega +{1\over 2\pi} g^2,
\end{equation}
for sufficiently large $r$. A crucial step in the derivation uses
of hypothesis $iii)$ to obtain $r^2\int_{S_r} \hbox{\rm (quadratic in }
p_{ab},\ \Dz_aq_{bc})\lambda^2 d\Omega \le 1/2\int_{S_r} \lambda^2 d\Omega$
for sufficiently large $r$.
Substituting (29) into (28), we obtain
\begin{equation}
r^2 \Bigl( {{dg}\over{dr}} \Bigr )^2 \le
r^2 (8 \pi +2r^2V(r))\int_{S_r} |{\cal D}_{AB} \mu_C|^2 d\Omega
+ {r^2 \over 2\pi}V(r)g^2.
\end{equation}
Dividing both sides of this last inequality by $(1+g^2)$,
the right side is integrable to $r=\infty$, and so therefore must
be the left side. Thus, just as in the proof of Theorem 2,
the function $g$ must approach a limit as $r$ approaches infinity,
and so we may subtract from $\lambda$ a constant to
make this limit zero. Having done so, we have that $r^2(dg/dr)^2$
is $r$-integrable (by Eqn. (30)), and so that $g^2$ is $r$-integrable
(by Eqn. (25)), and so that $\lambda^2/r^2$ is $T$-integrable
(by Eqn. (29)). The appropriate modifications of
the proof of Theorem 3 are similar but much simpler (requiring
only hypotheses $i)$ and $ii)$). \ \ /

Thus, for any initial-data set that is asymptotically flat
in the sense of the Theorem, the space \mcal\ has the
structure of a spinor space. In more detail, \mcal\ is a
complex, 2-dimensional vector space with an adjoint operation
$\dagger$ (obtained via conclusion $ii)$) and an alternating
tensor $\epsilon$ (obtained via conclusion $iii)$), with
these two having the usual properties: $\dagger \dagger=-1$;
$\epsilon(\alpha,\,\beta)=-\epsilon (\beta,\, \alpha)
=\overline {\epsilon(\alpha^\dagger,\, \beta^\dagger)}$ for
any $\alpha, \ \beta\in$ \mcal; and $\epsilon (\alpha, \alpha^\dagger)>0$
for any nonzero $\alpha\in$ \mcal. Think of \mcal\  as the space
of ``asymptotic spinors''. The space of ``asymptotic vectors''
is now obtained from \mcal\  by the usual construction
of vectors from spinors. Denote by ${\cal V}$ the collection
of all self-adjoint elements of the tensor product of
\mcal\
with its complex-conjugate space
$\overline {\cal S}$,
so ${\cal V}$ is a real, 4-dimensional vector space.
The alternating tensor $\epsilon$ on the spinor space \mcal\
gives rise to a Lorentz metric $g$ on the vector space
${\cal V}$; and then the adjoint operation $\dagger$ on the spinor
space \mcal\ gives rise to a unit timelike vector $t$
on the vector space ${\cal V}$. Think of $t$ as the ``asymptotic
normal to the surface $T$''.

We next introduce, from their definitions (17) and (18),
the two mass functions, $M_N$ and $M_A$, on the space \mcal.
It turns out (Theorem 6) that these two functions coincide
in this case. The mass function is a Hermitian quadratic form
on \mcal, and so gives rise to a linear function on ${\cal V}$,
i.e., to an element of the dual space, ${\cal V}^*$.
This covector on ${\cal V}$ is the total mass-momentum of
our initial-data set. Since the original quadratic form was
positive semi-definite, the mass-momentum is nonspacelike.
The $g$-norm of the mass-momentum is the invariant mass of
our initial-data set; its inner product with $t$, the
mass-component in the ``direction normal to the surface $T$''.
Thus, we have constructed the mass-momentum vector for any
initial-data set asymptotically flat in the  sense of
Theorem 4.

The asymptotic conditions of Theorem 4 are
weaker than those required in the standard$^1$ (ADM) definition
of mass-momentum.
Let us now impose on our initial-data set stronger asymptotic conditions:
that $r|q_{ab}-\qz_{ab}|,\  r^2|D_aq_{bc}|,$ and $r^2|p_{ab}|$
all be bounded. Now the ADM mass-momentum is well defined, and
so now comparison between it and the present mass-momentum
is possible. The two agree$^{14}$.
Thus, the present framework indeed represents a
generalization of the ADM mass-momentum.

Bartnik$^{15}$ has also generalized the ADM mass-momentum to weaker
a\-symp\-to\-tic conditions,
which appear to be very slightly stronger than ours.
Pre\-su\-mab\-ly, when both sets of conditions are satisfied,
the two mass-mo\-men\-ta agree.

\section{Properties}

We now  discuss some examples and some general properties
of the space \mcal\ and the two mass functions,
$M_N$ and $M_A$, thereon.

For certain initial-data sets, it can occur that \hhbar\ = \cbar,
i.e., that every spinor field $\lambda^A$ of finite norm (12)
is a limit of spinor fields of compact support. When this occurs,
we shall have the space \mcal\ zero-dimensional (with, e.g.,
representative $\lambda^A=0$), and so, necessarily, the
mass functions $M_N$ and $M_A$ vanishing.
This circumstance can arise in at least two ways ---
because there is ``too much matter'' in the space-time,
or ``too little room at infinity''. These are illustrated
in the following.

{\bf Theorem 5:}
Let $(T,\, q_{ab},\, p_{ab})$ be an initial-data set satisfying
the energy condition, and let $\kappa$ be a smooth positive
function on $T$
such that, for any number $\kappa_0$, the set
$B_{\kappa_0}=\{x\in T|\kappa(x)\le \kappa_0\}$
is compact. Further, let there exist a compact subset
of $T$ outside of which either

\qquad $i)\ \displaystyle \biggl
({D_a\kappa \over \kappa} \biggr )^2 \le \rho - |\rho^a|$
\qquad  everywhere, or

\qquad $ii)\ D^2\kappa +|p_{ab}D^b\kappa| \le 0$
\qquad  everywhere.

\noindent Then \mcal=$\{0\}$.

{\it Proof\/}:
Fix any $\lambda^A\in$\hcal, and any $\epsilon >0$. Choose $\kappa_0$
sufficiently large that, first, $B_{\kappa_0}$ contains
the compact subset of the theorem, and that, second,
\begin{equation}
\int_{(B_{\kappa_0})^c}
\Bigl [ |{\cal D}_{AB}\lambda_C|^2
+{1\over 2}
(\rho |\lambda|^2
+i\sqrt 2\rho_{AB}\lambda^{\dagger A}\lambda^B) \Bigr ] \le \epsilon.
\end{equation}
Let $f$ be the function, of compact support, given by
\begin{equation}
f= \cases{ 1,&$\kappa \le \kappa_0$\cr
1-s \log (\kappa/\kappa_0 ),&$\kappa_0< \kappa < \kappa_0e^{1/s}$\cr
0,&$\kappa_0e^{1/s}\le \kappa,$\cr}
\end{equation}
where $s>0$ is some number.
We now have
\begin{eqnarray}
\parallel \lambda -f\lambda \parallel ^2
&=& \int_T \Bigl \{ |\lambda|^2|Df|^2
+(1-f)^2 \Bigl [ |{\cal D}_{AB}\lambda_C|^2
+{1\over 2} (\rho |\lambda|^2
+i\sqrt 2\rho_{AB}\lambda^{\dagger A}\lambda^B) \Bigr ] \Bigr \}
\nonumber \\
&\le &\int_U{|\lambda|^2|Df|^2} + \epsilon,
\end{eqnarray}
where we have set $U=(B_{\kappa_0})^c\cap (B_{\kappa_0e^{1/s}})$.
In case $i)$, substitute $|Df|^2=s^2|D\kappa / \kappa |^2$ into
(33) and use condition $i)$ to obtain $\parallel \lambda -f\lambda \parallel^2
\le (2s^2+1)\epsilon$.
In case $ii)$, first choose $1>t>0$ sufficiently small that, setting
$w^a=t(D^a\kappa)/\kappa$, we have
$\int_{\partial B_{\kappa_0}}{|\lambda|^2 w\cdot dA } \le \epsilon$.
Next set $s^2=t(1-t)$, which, with condition $ii)$, yields
$D_a w^a + w^aw_a + |p_{ab}w^b| + |Df|^2 \le 0$ in $U$.
Then
\begin{eqnarray}
0 &\le &\int_{\partial (B_{\kappa_0e^{1/s}})}
{|\lambda|^2 w^a dA_a} \le \int_U{D_a(w^a|\lambda|^2)} +\epsilon
\nonumber \\
&=&\int_U \Bigl \{ |\lambda|^2D_aw^a
+i\sqrt 2w^{AB}p_{ABCD}\lambda^C\lambda^{\dagger D}
+w^{AB} \Bigl [\lambda^{\dagger C}{\cal D}_{AB}\lambda_C
-\lambda^C ({\cal D}_{AB}\lambda_C)^\dagger \Bigr ]\Bigr \} +\epsilon
\nonumber \\
&\le &\int_U \Bigl \{ |\lambda|^2(D_aw^a+|p_{ab}w^b|+w^2)
-|{\cal D}\lambda -w\lambda|^2 +|{\cal D}\lambda|^2 \Bigr \}
\ +\epsilon
\nonumber \\
&\le &\int_U{-|Df|^2 |\lambda|^2} +2\epsilon,
\end{eqnarray}
Now Eqn. (33) gives $\parallel
\lambda -f \lambda \parallel^2 \le 3\epsilon$.\ \ /

We give a number of applications of this theorem.

Let $(T,\,q_{ab},\, p_{ab})$ be an initial-data set
with $(T,\, q_{ab})$ complete.
Fix any origin $x$ on $T$ and let
$r$ denote $q$-distance from that origin.
Now suppose further that the data are such that,
for some positive number $c$,
$\rho-|\rho^a|\ge c^2/r^2$
outside a compact set.
Then \mcal$=\{0\}$.
( {\it Proof\/}: Set
$\kappa=(1+r^2)^{c/2}$
and apply Theorem 5, case $i)$.
This example renders in the present framework
the physical
statement
``when the mass density falls off more  slowly than
$1/r^2$, the total mass is infinite''.

As a second example, let $T$ be an open ball in $R^3$ with
radius $r_0$, let $q_{ab}$ the flat metric induced from that of $R^3$,
and let $p_{ab}=q_{ab}/(r_0-r)$. One can easily check that
here $\rho-|\rho^a|=1/(r_0-r)^2$, whence these
initial data satisfy the energy condition.
In this example,
\mcal$=\{0\}$. ({\it Proof\/}: Set $\kappa=(r_0-r)^{-1}$ and apply
Theorem 5, case $i)$.)

In these two examples of Theorem 5, case $i)$, there is
``too much matter in the space-time''$^{16}$.
The total physical mass of these initial-data sets is actually
``infinity''. But, since the present formalism permits
only finite values for the mass functions, $M_N$ and $M_A$,
the response of that formalism is to collapse
\mcal\ to a point, i.e., to provide no asymptotic directions along
which the components of the total mass-momentum can be evaluated.

For the next example, let $T=S^1\times S^1\times R$,\  $q_{ab}$ the
natural flat metric on this product, and $p_{ab}$ zero.
Thus, this is initial data for Minkowski space-time, but
with two dimensions ``suppressed by wrapping''. In this example,
we have \mcal$=\{0\}$. ({\it Proof\/}: Set $\kappa=r$ outside
a compact set, where $r$ denotes distance in $R$ from an origin,
and apply Theorem 5, case $ii)$.) Thus, the loss of
two ``asymptotic directions'' suffices to collapse \mcal\ to
a point. Indeed, even a single asymptotic direction suffices (but
just barely):
For $T=S^1\times R^2$, $q_{ab}$ again the natural flat metric and
$p_{ab}$ zero, we again have \mcal$=\{0\}$. ({\it Proof\/}:
Set $\kappa=\log r$ outside a compact set, where
$r$ denotes distance in $R^2$ from an origin, and again
apply Theorem 5, case $ii)$.)

In general, according to case $ii)$ of Theorem 5, the space \mcal\
collapse to a point whenever there is ``too little room at infinity''.
This is illustrated in the examples of the previous paragraph,
in which entire asymptotic dimensions are suppressed. A second
class of examples involves ``nearby'' asymptotic regions.
Let $(T,\, q_{ab},\, p_{ab})$ be any initial-data set with $T$ compact.
Then clearly \ccal\ =\hcal, and so we have
\mcal$=\{0\}$.
Let us
now remove a single point $x$ from $T$. The resulting
initial-data set again has
\mcal$=\{0\}$.
({\it Proof\/}: Set $\kappa=r^{-1/2}$ in a neighborhood of $x$, where
$r$ denotes $q$-distance from $x$, and apply Theorem 5, case $ii)$.)
Similarly, the removal of any finite number of points from an
initially compact initial-data set retains \mcal$=\{0\}$.
Next, let us remove from the compact $T$ above any
closed curve. Then, again, we have
\mcal$=\{0\}$.
({\it Proof\/}: Set $\kappa=(-\log r)^{1/2}$ in a neighborhood of
that curve, where $r$ denotes $q$-distance from the curve,
and apply Theorem 5, case $ii)$.)
Similarly, the removal of any finite number of closed
curves and line segments retains
\mcal$=\{0\}$.

The examples of the previous paragraph show that an otherwise
compact initial-data set with a zero- or one-dimensional ``edge''
behaves as though it were compact: It retains \mcal=$\{0\}$
and, therefore, zero mass functions. What happens in the
case of a two-dimensional ``edge''? It turns out that, in this case,
the space \mcal\ is always infinite-dimensional, and the mass
functions $M_N$ and $M_A$ are always nontrivial. This we now show.
Let $\overline T$ be a smooth, compact three-dimensional manifold
with boundary, so $K=\partial \overline T$ is a smooth, two-dimensional
manifold. Fix smooth initial data, satisfying the energy condition,
on $\overline T$. Now consider the initial-data set $(T,\, q_{ab},\, p_{ab})$,
where $T={\overline T}- K$ is the interior of $\overline T$,
and $q_{ab}$ and $p_{ab}$ are the induced initial data.
We wish to determine the space \mcal\ and the mass functions
$M_N$ and $M_A$ for this initial-data set.

Consider first any smooth spinor field $\lambda^A$ on $\overline T$.
Then its restriction to $T$ certainly defines an element of \hcal.
When are two such equivalent, i.e., when do they differ by
an element of \cbar ? We claim: $\lambda^{'A}-\lambda^A \in$\cbar\
if and only if $\lambda^{'A}=\lambda^A$ on the boundary $K$.
To prove ``if'', assume that $\lambda^{'A}=\lambda^A$ on $K$,
and set, for $n=1,2,\ldots ,$
\begin{equation}
\lambda_n^A= \cases{0,&$r\le 1/n$ \cr
(nr-1)(\lambda^{'A}-\lambda^A ), & $1/n < r < 2/n$ \cr
\lambda^{'A}-\lambda^A, & $2/n \le r$ }
\end{equation}
where $r$ is $q$-distance from $K$. Then verify that this
sequence of spinor fields is in \ccal, and that it converges
in \hhbar\ to $\lambda^{'A}-\lambda^A$. To prove ``only if'',
first note that, for any fields $\mu^A$ and $w^a_{\ B}$ on
$\overline T$, we have
\begin{eqnarray}
\int_K w^a_{\ B} \mu^B dS_a &=& \int_T {\cal D}_a(w^a_{\ B} \mu^B)
= \int_T ({\cal D}_aw^a_{\ B}) \mu^B +\int_T w^a_{\ B}({\cal D}_a \mu^B)
\nonumber \\
&\le & \biggl [ \int_T |D_aw^a_{\ B}|^2 \int_T |\mu|^2 \biggr ]^{1/2}
+ \biggl [ \int_T |w^a_{\ B}|^2 \int_T |{\cal D}_a \mu^B |^2
\biggr ]^{1/2}.
\end{eqnarray}
But each term on the right is continuous in $\mu^B\in$\hcal\
(using Theorem 1 for the first term), and so the surface integral
on the left is
also continuous. The conclusion --- that $\lambda_n^A\in$\ccal\
converging in \hhbar\ to $\lambda^{'A}-\lambda^A$ implies
$\lambda^{'A}=\lambda^A$ on $K$ --- follows.

Thus, each smooth spinor field specified on $K$ gives rise to
a point of \mcal, and distinct such fields give rise
to distinct points of \mcal. So, in particular, the space \mcal\
in this example is infinite-dimensional. It is easy to evaluate
the asymptotic mass function at such points of \mcal. Indeed,
for $\lambda^A$ a spinor field specified on $K$, giving rise
to point $\alpha$ of \mcal, we have$^3$, from Eqn. (11),
\begin{equation}
M_A(\alpha)=\int_K (\lambda^{\dagger C}{\cal D}_{AB}\lambda_C
+2{\lambda^\dagger}{_{(B}}{\cal D}^{\ }_{A)C}\lambda^C )dS^{AB}.
\end{equation}
Note that the right side does indeed depend only on
$\lambda^A$ at points of $K$, i.e., it only involves derivatives
of $\lambda^A$ tangent to $K$. Thus, the asymptotic mass function,
at such points, is a simple surface integral. The
norm mass function is of course more complicated.

The full space \mcal\ in this example is constructed essentially by
``completing'' the collection of special elements obtained above.
The result$^{17}$ is the Sobolev space $W^{1/2}(K)$.
This consists, in more detail, of those spinor fields specified
on $K$ such that the integral
\begin{equation}
\int_K |\lambda|^2
+\int_K dS_x
\int_K dS_y
{{|\lambda(x) -\lambda(y)|^2}\over{d(x,y)^3}}
\end{equation}
(which defines the Sobolev norm) converges. Here, $d(x,y)$ denotes
the $q$-distance between points $x$ and $y$ of $K$, and the
difference of spinors in the second integrand is to be taken
using components in any basis. The spinor fields in $W^{1/2}(K)$
are better behaved than those in $W^0(K)=L^2(K)$, but not
so well-behaved as those in $W^1(K)$ (whose norm is given
by the integral over $K$ of the square of the spinor field
plus the square of
its derivative). The integral on the right in Eqn. (37)
makes sense for spinor fields $\lambda^A$ in $W^{1/2}(K)$,
and again yields the asymptotic mass function $M_A$.

Thus, the space \mcal\ in this example is infinite-dimensional,
and the mass functions, $M_N$ and $M_A$, are complicated bounded,
quadratic functions on \mcal. Even in the example of a ball
from Minkowski initial data, there are points of \mcal\ at which
$M_N$ is positive (and so we shall not have $M_N=0$), and
points at which $M_A$ is negative (and so we shall not have $M_A=M_N$).

There is a simple inequality relating the two mass functions
to each other:
\begin{equation}
-2M_N(\alpha)\le M_A(\alpha)\le M_N(\alpha).
\end{equation}
To derive this, take the infimum of the right side of (18)
and use $|{\cal D}_{AB}\lambda^B|^2$ $\le {3\over 2}
|{\cal D}_{AB}\lambda_C|^2$.
There are, as we shall see shortly, many examples for
which the second inequality in (39) is an equality. Are
there nontrivial examples for which the first inequality
is an equality?

For which initial-data sets must
the two mass functions be equal?
There is, as it turns out, a large class for which
we can assert that $M_A=M_N$, these consisting of
``essentially all'' complete initial-data sets.

{\bf Theorem 6:}  Let $(T,\,q_{ab},\,p_{ab})$ be
an initial-data set satisfying the energy condition.
Assume that

$i)\ T,\,q_{ab}$ is complete, and

$ii)$ $T$ admits no nonzero spinor field $\kappa^A\in L^2(T)$
with $\parallel\kappa\parallel^2=0$.

\noindent Then $M_A=M_N$.

{\it Proof\/}: Fix any $\alpha\in$ \mcal, and let $\lambda^A$
be any representative. So ${\cal D}_A^{\ \ B}\lambda_B\in L^2(T)$.
Denote by $\kappa^A$ the $L^2$-projection of
${\cal D}_A^{\ \ B}\lambda_B$ orthogonal to the closed
subspace of $L^2$ generated by elements of the form
${\cal D}_A^{\ \ B}\sigma_B$ with $\sigma\in$\ccal.
Then
 $\kappa\in L^2(T)$, and (by orthogonality to the
${\cal D}_A^{\ \ B}\sigma_B$)
${\cal D}_A^{\ \ B}\kappa_B=0$. We have only to show
that $\kappa^A=0$, for this implies, by Eqns. (17) and (18),
that $M_A(\alpha)=M_N(\alpha)$.

For $f$ any function on $T$ of compact support, we have
\begin{eqnarray}
&&\int f^2\bigl[|{\cal D}_{AB}\kappa_C|^2+{1\over 2}
(\rho |\kappa^A|^2+i\sqrt2 \rho_{AB} \kappa^{\dagger A}\kappa^B)\bigr]
\nonumber \\
&=&\int f^2 {\cal D}^{AB}(\kappa^{\dagger C}{\cal D}_{AB}\kappa_C)
\nonumber \\
&\le &\biggl [\int |Df|^2|\kappa|^2 \biggr ]^{1/2}
\biggl [ \int f^2 \bigl[|{\cal D}_{AB}\kappa_C|^2+{1\over 2}
(\rho |\kappa^A|^2+i\sqrt2 \rho_{AB} \kappa^{\dagger A}\kappa^B)\bigr]
\biggr ]^{1/2},
\end{eqnarray}
where we used (11) in the first step, and an integration by parts
and the Schwarz inequality in the second. Now let $r$ denote
$q$-distance in $T$ from some fixed point, and set, in Eqn. (40),
\begin{equation}
f=\cases{ 1,&$r\le r_0$\cr
2-r/r_0,&$r_0<r<2r_0$\cr
0,&$2r_0\le r$\cr}
\end{equation}
where $r_0$ is some number. This $f$ has compact support,
by hypothesis $i)$. Letting $r_0$ approach infinity, the
left side of Eqn. (40) approaches $\parallel\!\kappa\!\parallel^2$,
while, since $\kappa\in L^2$ and $|Df|\le 1/r_0$, the first factor
on the right approaches zero. So, $\parallel\!\kappa\!\parallel^2=0$,
and so, by hypothesis $ii)$, $\kappa^A=0$.\ /

Hypothesis $ii)$ of Theorem 6 serves only to rule out a few,
very special, examples. Indeed, as we shall see shortly,
the only initial-data sets admitting a nonzero $\kappa^A$
with
$\parallel\!\kappa\!\parallel^2=0$
are certain ones for flat space-times and certain ones for plane-wave
space-times. So, for instance, hypothesis $ii)$ can be dropped
entirely for any initial-data set having some point
at which $\rho>|\rho^a|$, or some point
at which the Weyl tensor is other than type [-] or type [4].

But, nevertheless, Theorem 6 is actually false without
hypothesis $ii)$. For example, fix constant spinor fields
$\kappa^A$ and $\mu^A$ in Minkowski space-time, with these
normalized by $\kappa^A\mu_A=1$. Set
$\lambda^A=\kappa^A\mu_B {\overline \kappa}_{B'}x^{BB'}$,
where $x^{BB'}$ is a dilation vector field
(i.e., one satisfying $\nabla_ax^b=\delta_a^{\ b}$).
Then, for any slice
$T$ in this space-time, $\kappa^A$ and $\lambda^A$ become spinor
fields on the corresponding initial-data set. Choosing
the slice such that its unit normal, $t^{AA'}/\sqrt2$, lies
in the plane of $\kappa^A{\overline \kappa}^{A'}$
and $\mu^A{\overline \mu}^{A'}$, we have
\begin{equation}
{\cal D}_{AB}\kappa_C=0, \quad |\kappa|^2=|{\cal D}_{AB}\lambda_C|^2
=i\kappa^A{\cal D}_A^{\ \ B}\lambda_B=t_{AA'}\kappa^A{\overline \kappa}^{A'}.
\end{equation}

Now choose Minkowskian coordinates, $(t,x,y,z)$, such that
$\kappa^A{\overline \kappa}^{A'}$ and $\mu^A{\overline \mu}^{A'}$ have
respective components (1,1,0,0) and (1,-1,0,0), let
$T$ be given by $t=x(1-(1+x^2)^{-1/2})$, and cyclically
identify $y$ and $z$ (i.e., identify points
$(t,x,y,z)$ and $(t,x,y+n,z+m)$, for $n,\ m$ any integers).
The resulting initial-data set has spinor fields $\kappa^A$ and
$\lambda^A$ with (by Eqn. (42)) ${\cal D}_{AB}\kappa_C=0,\ \ \kappa\in L^2,\
{\cal D}_{AB}\lambda_C\in L^2,\ \int \kappa^A{\cal D}_A^{\ \ B}\lambda_B
\neq 0$. It follows from the third of these that $\lambda\in$\hcal,
and so that this $\lambda^A$ defines some element, $\alpha$, of
${\cal S}$. But the other three properties imply that
${\cal D}_A^{\ \ B}\lambda_B$ cannot be made arbitrarily
small (in $L^2$) by addition to $\lambda^A$ of various $\sigma^A\in$\ccal.
This shows that $M_A(\alpha)\neq M_N(\alpha)$, and, in particular, that
$\alpha\neq 0$.

For incomplete initial data, there are much simpler examples
in which $M_A$ and $M_N$ differ.
For instance, fix $r_0>0$ and
let $(T, q_{ab},\, p_{ab})$
be the initial-data set obtained from the subset
$r_0<r<4r_0$ of the standard flat initial-data set,
where $r$ is
$q$-distance from some fixed origin.
Choose spinor field $\lambda^A$ on $(T,\, q_{ab})$
be a constant
field, $\lambda^{(1)}_A$ on $r_0<r<2r_0$,
and a different constant field, $\lambda^{(2)}_A$ on
$3r_0<r<4r_0$.
Denoting by $\alpha\in$\mcal\ the corresponding equivalence
class, it is easy to check that
$M_A(\alpha)=0$ and $M_N(\alpha)={{16\pi}\over
{3}}r_0|\lambda^{(1)}_A-\lambda^{(2)}_A|^2$.

We now turn to the issue of the positivity of
the two mass functions.
The norm mass function, from its definition
in (17), is manifestly
nonnegative,
while
the asymptotic mass function,
from its definition (18), is not.
When the data are complete,
the two mass functions are generally the same,
and so nonnegativity also of the asymptotic mass function then follows.

Can the asymptotic mass function ever become negative?
The answer is yes, as we see in the following example.
Let $(T,\, q_{ab},\, p_{ab})$ be the initial-data set with
$T$ any open subset, with compact closure, in $R^3$,
$q_{ab}$ the metric on $T$ inherited from the
Euclidean metric of $R^3$,
and $p_{ab}=0$.
Set
\begin{equation}
\lambda_A=\beta^\dagger_A
\beta_M \beta_N \beta_P \beta_Q
x^{MN}x^{PQ},
\end{equation}
where
$x^a$ is a dilation vector field on $T$,
and $\beta_A$ any nonzero constant spinor field, normalized by
$\beta^{\dagger A}\beta_A=1$.
Then for any spinor field $\sigma_A$
on $T$ of compact support we have
\begin{equation}
\int_T{|({\cal D}_{AB}(\lambda_C+\sigma_C)|^2}
=\int_T
{|{\cal D}_{AB}\lambda_C|^2
+|{\cal D}_{AB}\sigma_C|^2},
\end{equation}
for the cross term vanishes by virtue of
${\cal D}^{\dagger A}_{\ \ B}
{\cal D}^B_{\ \,C}\lambda^C=0$.
Hence, this $\lambda_A$
realizes the infimum in (18),
i.e., we have
\begin{equation}
M_N(\alpha)=\int_T{
|{\cal D}_{AB}\lambda_C|^2}
=\int_T{|\beta_A\beta_Bx^{AB}|^2},
\end{equation}
where $\alpha\in$\mcal\ is the corresponding equivalence class.
Substituting into (18), we now obtain
$M_A(\alpha) =-M_N(\alpha)<0$.
Is there a simple theorem guaranteeing $M_A\ge 0$ for some
large class of initial-data sets?

We have seen that the norm mass function, by
its definition, can never be strictly negative.
But it is sometimes possible
for this mass function
to attain the value zero,
e.g., in the case of data for Minkowski space-time.
When, more generally, can this occur?
Fix an initial-data set $(T,\, q_{ab},\, p_{ab})$
satisfying the energy condition,
and denote by $Z$ the collection of all
spinor fields $\lambda^A$ on $T$ that are
${\cal D}$-constant:
\begin{equation}
{\cal D}_{AB}\lambda_C=0.
\end{equation}
Then $Z$ is clearly a complex vector space, with
(since two solutions of (46) agreeing at a point
must agree everywhere) dimension at most two.
By (11), $\lambda^A\in Z$\ if and only if
$\parallel\!\lambda\!\parallel^2=0$.
Thus, each element of $Z$ gives rise to a
point$^{18}$
of \mcal\ at which $M_N=0$.
Are these the only points of \mcal\ at which the norm
mass function vanishes? That they are (i.e., that
$\alpha\in$\mcal\ with $M_N(\alpha)=0$ implies
that there is a representative $\lambda^A$ of $\alpha$
with $\lambda^A\in Z$) would follow from:

{\bf Conjecture 7.}  Let $(T,\ q_{ab},\ p_{ab})$
be an initial-data
set satisfying the energy condition, and $x$ any point of $T$.
Then there exists a neighborhood $U$ of $x$ and a number $c>0$
with the following property: Given any $\lambda^A\in$\hcal,
there is a field
\lambdaz$^A$
in $U$, there satisfying
$(46)$, such that
\begin{equation}
\int_U{ \Bigl [ |{\cal D}^{AB}\lambda^C|^2 +{1\over 2} (\rho\lambda^{\dagger A}
\lambda_A +i\sqrt 2\rho_{AB}\lambda^{\dagger A}\lambda^B)} \Bigr ]
\ge c\int_U{|\lambda - \lambdaz |^2}.
\end{equation}
This conjecture$^{19}$
asserts, roughly speaking, that, locally and
modulo fields of norm zero, the norm $\parallel\cdot\parallel^2$
bounds the $L^2$ norm.
Note, e.g., that the conclusion of the conjecture
holds automatically whenever
$\rho > |\rho^a|$\  at the point $x$.
There follows from this conjecture,
not only that all zeros of $M_N$ arise from
$Z$, but an even
stronger result, to the effect that
all ``near zeros'' of $M_N$ also arise from $Z$.
More precisely, we have the following consequence
of Conjecture 7:
For $\alpha_1,\ \alpha_2,\ \ldots$
points of \mcal, with $M_N(\alpha_i)$ approaching zero,
there exist elements $\mu_1^A,\ \mu_2^A,\ \ldots$
of $Z$ such that the sequence $\alpha_i-\{\mu_i\}$
in \mcal\ approaches zero.
This consequence would guarantee, e.g., that, whenever
$Z=\{0\}$, \mcal\ is a Hilbert space under norm $M_N$.
Thus, Conjecture 7 would
provide good control, in terms of the simple
vector space $Z$, of all the ``zero behavior''
of the norm mass function.
It would be of interest to settle this conjecture.

To see what Eqn. (46) means geometrically,
we proceed as follows. Fix an initial-data set
satisfying the energy condition, and admitting
a nonzero solution, $\lambda^A$, of Eqn. (46).
It is convenient to embed this initial-data set in a
full space-time satisfying the dominant energy condition:
that $(R_{ab}-1/2Rg_{ab})u^b$ is future-directed nonspacelike
for every future-directed timelike $u$.
(Such an embedding is always possible,
e.g., by choosing for the matter source dust.)
Thus, we obtain a full, 4-dimensional space-time,
$(M,\, g_{ab})$, with a certain spacelike, 3-dimensional
submanifold $T$. Then $\lambda^A$, the spinor field
defined originally on the manifold $T$, becomes
a spinor field in the full space-time, defined only
at points of the submanifold $T$. Eqn. (46) becomes that
\begin{equation}
w^b\nabla_b\lambda^A=0
\end{equation}
at each point of the submanifold $T$, where
$w^a$ is any vector at that point tangent to $T$.
Taking a second derivative tangent to $T$ and
commuting, we find
\begin{equation}
t_{[a}R_{bc]de}l^e=0, \qquad t_{[a}R_{bc][de}l_{f]}=0,
\end{equation}
at all points of $T$, where $t^a$ is the unit normal to $T$,
and $l^a$ the null-vector equivalent of the spinor $\lambda^A$.
Contraction the second of these equations three times, we obtain
\begin{equation}
(R_{ab}-{1\over 2} Rg_{ab})l^at^b=0
\end{equation}
But this, along with the dominant energy condition, implies
in turn that $R_{ab}$ is some multiple of $l_al_b$,
i.e., that the matter is null dust.
Substituting this Ricci tensor into Eqns (49), we
obtain these same equations on the Weyl tensor, $C_{abcd}$.
But these in turn imply
\begin{equation}
C_{bcde}l^e=0, \qquad C_{bc[de}l_{f]}=0,
\end{equation}
i.e., that the Weyl tensor is type [4], with $l^a$
the repeated principal null direction.

To summarize, in the ``generic'' case, the vector space $Z$
is zero-di\-men\-sion\-al, and the norm mass function is strictly
positive on nonzero elements of \mcal. In order that $Z$
be one-dimensional, the matter must be  null dust,
and the Weyl tensor type [4] with the dust velocity as
its principal null
direction. In order that $Z$ be two-dimensional, the initial-data
must be flat. The space $Z$ can never be of dimension three
or higher.

What happens under change in the initial data
in some compact region? Will the space \mcal\ and the mass functions
also change?
To address this issue,
consider two initial-data sets,
$(T,q_{ab},p_{ab})$\ and $(T,\widetilde q_{ab},\widetilde p_{ab})$,
on the same underlying manifold $T$. Let both satisfy
the energy condition, and
let $q_{ab}=\widetilde q_{ab}$\ and $p_{ab}=\widetilde p_{ab}$
outside some open subset $U$
of $T$ with compact closure.
We first obtain
a natural isomorphism $\Im$ between the corresponding
\mcal\ and $\widetilde {\cal S}$, as follows.
Given any element of \mcal,
with representative $\lambda_A$,
map it via $\Im$ to
that element of ${\widetilde {\cal S}}$
having some representative,
$\widetilde \lambda_A$,
with  $\widetilde \lambda_A=\lambda_A$
outside $U$.
This makes sense,
since $q_{ab}=\widetilde q_{ab}$
and
$p_{ab}=\widetilde p_{ab}$
outside $U$; and is independent of representative,
since $U$ has compact closure.
This map is indeed a continuous isomorphism
between \mcal\ and $\widetilde {\cal S}$.

We first note that the asymptotic
mass function $M_A$ is invariant
under this isomorphism, for it
can be written (by Eqns. (18) and (11)) as an
integral whose integrand is a pure divergence.
But what of the norm mass function?
For complete initial data,
the norm mass function is usually the same as the asymptotic mass
function and therefore must also be invariant under
$\Im$.
But, for incomplete data, invariance can fail.
For example, let $T$ be any open, bounded subset of
$R^3$, $q_{ab}$ the flat metric on $T$ induced from
that of $R^3$, and
$p_{ab}=2\sqrt 2/3 s_0 q_{ab}$,
where $s_0$ is any positive constant.
This initial-data set satisfies
the energy condition (indeed, with $\rho={8\over 3}s_0^2$, $\rho_a=0$).
Set
\begin{equation}
\lambda^A=\exp (ik_ax^a) {\lambdaz}{^A}
\end{equation}
where
\lambdaz$^A$ is any constant spinor field, normalized by
$|\lambdaz|^2=1$,
$k_a$ is the constant vector field given by
$k_{AB}=2is_0$ \lambdaz$^\dagger {_{(A}}\lambdaz_{B)}$, and
$x^a$ is a dilation vector field.
This $\lambda^A$ is in \hcal,
and so defines an element, $\alpha$, of \mcal.
Using that
${\cal D}^{\dagger A}_{\ \ \, B}
{\cal D}^B_{\ \,C}\lambda^C=0$,
it follows, from Eqn. (17) and (11) that
\begin{equation}
M_N(\alpha)=\parallel\!\lambda\!\parallel^2.
\end{equation}
Now change this initial data, to
$\widetilde q_{ab} = q_{ab}$\
and
$\widetilde p_{ab}= p_{ab}-2\sqrt 2/3\ h q_{ab}$,
where $h$ is a smooth function on $T$ of compact support.
For $h$ and its derivative sufficiently small, these will continue to
satisfy the energy condition.
Then from Eqn. (53) and the fact that $M_N$ is
defined as an infimum, we have
\begin{eqnarray}
M_N(\widetilde \alpha) &\le & M_N(\alpha)
+2\int { \Bigl [ ih \bigl (\lambda^{\dagger A}{\cal D}_{AB}\lambda^B-
\lambda^A({\cal D}_{AB}\lambda^B)^\dagger\bigr )
+h^2 \lambda^{\dagger A}\lambda_A \Bigr ]}
\nonumber \\
&=& M_N(\alpha)+2\int{(h^2-2s_0h)|\lambda|^2}
\end{eqnarray}
Now choosing $h$ non-negative and sufficiently small,
the last integral on the right becomes negative, yielding
$M_N(\widetilde \alpha)
<  M_N(\alpha)$. Thus, a change in the data,
though restricted to a compact region, has nontheless
changed the norm mass function.
Is there any simple
theorem that isolates a large class of initial-data sets
having invariance of $M_N$ under compactly-supported changes
in the data?

Some initial-data sets contain
several ``asymptotic regions''.
A familiar example is that of a slice
in the extended Schwarzschild spacetime that extends through the
``wormhole-throat''.
We consider now some properties of
\mcal\ and the mass functions in such examples.
Let
$(T,\, q_{ab},\,p_{ab})$
be an initial-data set satisfying the energy
condition.
Fix closed subsets, $T_1$ and $T_2$, of $T$
that have compact intersection and cover $T$.
These represent the two ``asymptotic regions''.
We first show that, under this arrangement, the
space \mcal\ splits as a direct sum of two subspaces.
Denote by \hcal$_1$ the collection of all spinor fields
$\lambda^A\in$\hcal\ having supp($\lambda)\subset\ T_1\cup A$
for some compact set $A$, and similarly for \hcal$_2$.
Then each of \hcala\  and \hcalb\  is a vector subspace
of \hcal, while these two
have intersection \ccal\ and together span \hcal.
Furthermore, we have that any field $\lambda^A$ in
\hhbar$_1\cap$\hhbar$_2$ is also in \cbar.
(Indeed, for
$\lambda_i^{(1)}\in$\hcala\ and
$\lambda_i^{(2)}\in$\hcalb,
each converging to $\lambda$, we have,
choosing any smooth function $f$ with
supp$(f)\cap T_1$ and
and supp$(1-f)\cap T_2$
both compact, that the
$f\lambda_i^{(1)} +
(1-f)\lambda_i^{(2)}\in$\ccal\
converge in \hhbar\ to $\lambda$.)
But these facts together imply that the space
\mcal\ is the direct sum of its two subspaces,
\mcal$_1$=\hhbar$_1$/\cbar\ and \mcal$_2$=\hhbar$_2$/\cbar.
In the example of the Schwarzschild space-time, each of
\mcal$_1$ and \mcal$_2$ becomes the two-dimensional vector
space of ``asymptotic spinors'' in the appropriate asymptotic
region.

Denote by $M_{1A}$ and $M_{2A}$ the restrictions of the
asymptotic mass function $M_A$ to the respective subspaces
\mcal$_1$ and \mcal$_2$ of \mcal.
Similarly, $M_{1N}$ and $M_{2N}$ for the norm mass function.
Thus, we introduce, for each of the two asymptotic regions,
separate mass functions. In the case of the asymptotic
mass, these separate mass functions add to give the total
mass function, i.e., we have$^{20}$
$M_A=M_{1A}+M_{2A}$. (This
is easily seen by noting that every $\alpha\in$\mcal\
has a representative of the form $\lambda_{1A} + \lambda_{2A}$
with $\lambda_{1A}\in$\hcala\ and $\lambda_{2A}\in$\hcalb;
and recalling that the right side of (18) is independent
of representative.) It follows from this that also
$M_N=M_{1N}+M_{2N}$ provided the initial-data set
is one in which
$M_N=M_A$, e.g., is one to which Theorem 6 applies.
For example, in the case of the slice in the extended
Schwarzschild space-time, $M_{1N}$ and $M_{2N}$ yield
the usual mass-momenta corresponding to the
respective asymptotic regions.
Theorem 6 applies to this example, and so
the total norm mass function is just the sum of these two.

We remark, however, that in general  we need not have
$M_N=M_{1N}+M_{2N}$. For example, fix numbers $0<r_1<r_2$,
let $(T,\, q_{ab},\, p_{ab})$ be the subset $r_1<r<r_2$ of
the standard initial data for Minkowski space-time,
where $r$ is distance from some origin. Let $T_1$ and
$T_2$ denote the subsets given by
$r_1<r\le {1\over 2}(r_1+r_2)$
and
${1\over 2}(r_1+r_2)\le r< r_2$, respectively.
Fix a constant spinor field \lambdaz$^A$ on $(T,\, q_{ab})$,
and let $\alpha\in$\mcal\ denote its equivalence class.
Then, by direct calculation, one verifies that
$M_A(\alpha)=M_{1A}(\alpha)=M_{2A}(\alpha)=M_N(\alpha)=0$,
while $M_{1N}(\alpha)=M_{2N}(\alpha)=
{\displaystyle {{4\pi r_1 r_2}\over
{r_2-r_1}} }
|\lambdaz|^2$.

It follows in particular from the observations above, and
the examples at the beginning of this section, that the
removal of any finite number of points, closed curves, and
closed line segments from an initial-data set changes neither
the space \mcal\ nor the mass functions $M_N$ and $M_A$.

All the remarks above can be generalized, easily, to any
finite number of asymptotic regions, and, somewhat less
easily, to any infinite number.

\section{Conclusion}

We have constructed, for any initial-data set
$(T,\, q_{ab},\, p_{ab})$ satisfying the energy condition,
a complex vector space, \mcal, representing ``asymptotic spinors'',
and two functions, $M_N$ and $M_A$, on \mcal, representing
``components of total asymptotic mass-momentum''.
For an asymptotically flat initial-data set, this framework
reproduces the standard mass-momentum at spatial infinity.
We have derived a number of general properties of these objects,
and applied this construction to a number of examples.
There follows a discussion of some open questions and
outstanding issues.

Fix a space-time that is asymptotically flat at null infinity
$^{21}$
and con\-si\-der space\-like slices, approaching cross-sections of
null infinity, in this space-time. The present formalism can,
of course, be applied to the initial-data sets so constructed.
Can we thereby recover$^{22}$ Bondi mass-momentum$^{21}$
at null infinity? Consider first the flat case. Let
$(T,\, q_{ab}, \, p_{ab})$ be the initial-data set that
arises from the slice $T$ in Minkowski space-time given by
the hyperboloid of points unit timelike distance from some
origin. What is \mcal\ for this example$^{23}$?
Each nonzero constant spinor field on the Minkowski space-time gives
rise to a ${\cal D}$-constant field on $T$, and so
to an element of the space \mcal. The elements of \mcal\
so constructed are, presumably, nonzero, and so we obtain
a 2-dimensional subspace of \mcal. Does this subspace
exhaust \mcal? Assuming that it does, we immediately
acquire an alternating tensor on this \mcal,
for the inner product of two spinor fields so constructed
is constant. But note that we do not acquire an adjoint
operation on \mcal, for the adjoint of a spinor field
in \hcal\ in this example is not in general in \hcal.
Does there exist a result, similar to Theorem 4, asserting
that any initial data that is ``asymptotically hyperbolic''
in an appropriate sense produces a space \mcal\ of
similar structure? Does the mass function $M_N$ now
reproduce the Bondi mass-momentum at null infinity?
Can this be generalized to include slices
that approach cross-sections of null infinity in ways
different from those above?

The present framework is intended to describe total
asymptotic mass-momentum. Is there anything analogous for
angular momentum? It seems likely that, if there is, then
there will be needed a new space to replace \mcal.
Indeed, in the asymptotically flat case, angular momentum,
because of its origin-dependent character, cannot be
expressed as any structure over the space \mcal\ of
asymptotic spinors.

Let $(T,\,q_{ab},\,p_{ab})$ and $(T,\,\widetilde q_{ab},\,\widetilde p_{ab})$
be two initial-data sets, based on the same underlying manifold $T$,
with each satisfying the energy condition. Under what conditions
can we construct a natural correspondence between their spaces
\mcal\ and $\widetilde {\cal S}$? We have seen in Sect. 3 that
there is such a correspondence when the data are identical
outside some compact subset of $T$. Furthermore, Theorem 4
can be interpreted as providing just such a correspondence
under the assumptions that $T=R^3$, $q_{ab}$ is Euclidean,
$p_{ab}$ is zero, and $(\widetilde q_{ab},\,\widetilde p_{ab})$
approaches $(q_{ab},\,p_{ab})$ sufficiently rapidly as $r$
approaches infinity. Is there a generalization of these
observations? Is there some simple theorem guaranteeing
that \mcal=$\widetilde {\cal S}$ for general initial-data sets
approaching each other asymptotically at an appropriate rate?

How much of the present formalism survives when there is
no longer imposed the energy condition, Eqn. (3)? One
might expect that everything will go through as before,
with the sole exception that now the mass function $M_N$
can become negative. But it turns out that, in the absence of the
energy condition, the entire formalism disintegrates.
We originally defined the space \hcal\ as consisting of
those spinor fields for which the
integral on the right in Eqn. (12) converges. But in the
absence of the energy condition we can no longer guarantee nonnegativity
of the integrand: What, then, is ``converge'' to mean? One
could, for example, require absolute convergence of the entire
integral, or absolute convergence of the integral of each term.
But neither of these, as it turns out, results in general in an
\hcal\ even having the structure of a vector space! It is possible
to recover a vector-space structure for \hcal, e.g., by
requiring convergence of the integrals of each of the first two
terms in (12), and also of $|\rho_a||\lambda|^2$ (which guarantees absolute
convergence of the integral of the last term). But this version
also appears to be unsatisfactory, for, when the energy condition {\it is}
satisfied, the space \hcal\ it produces is in some
cases strictly smaller than the space \hcal\ as originally defined.

What is \mcal\ for the general initial data for flat space-time?
That this question may not be as simple as it appears is suggested
by the following examples. We introduced at the beginning of this
section the example of the hyperboloid in Minkowski space-time
consisting of points unit timelike distance from some origin.
While it appears likely that \mcal\ in this case is 2-dimensional,
corresponding
to the constant spinor fields in Minkowski space-time, a proof
in lacking. A more complicated example is that following Theorem 6
of Sect. 3. Again, constant spinor fields on the Minkowski
space-time give rise to ${\cal D}$-constant spinor fields,
$\mu^A$ and $\kappa^A$,
on $T$, and so to points of \mcal.
The point of \mcal\ associated
with the field $\kappa^A$ turns out to be zero (by the proof of Theorem 6),
but we acquire a new point of \mcal\ from the spinor field $\lambda^A$ given in
that example. So, we end up with a 2-dimensional subspace
of \mcal\ --- associated with the elements $\mu^A$ and $\lambda^A$. Does this
subspace exhaust \mcal? Another example is that of a cosmic string:
{}From the initial-data set with $(T,\,q_{ab})$ Euclidean space and
$p_{ab}$ zero,
remove a straight line and introduce a deficit angle. What is \mcal\
for this example? We suggest that the most likely answer is that
\mcal=$\{0\}$. Let $\lambda^A\in$ \hcal. Then Theorem 5, case $ii)$
suggests that $\lambda^A$ can be approximated near the singular
axis by fields of compact support. (Were the singular axis
compact, then the theorem would apply directly.) Furthermore,
Theorem 2 can be modified to show that $\lambda^A$ approaches a constant
asymptotically within ``any fixed, small solid angle''.
But, because of the presence of the deficit angle,
the only asymptotically constant spinor field in this example
is zero. This suggests that $\lambda^A$ approaches zero asymptotically,
and so can be approximated in the asymptotic region by fields of
compact support. The above is only a plausibility argument that
every $\lambda^A\in$ \hcal\  can  be approximated in \hcal\ by
spinor fields of compact support, and so \mcal$=\{0\}$.
We do not know for sure what the space
\mcal\ is in this example. This last example, by virtue of its
deficit angle, carries no ${\cal D}$-constant spinors.
There can also be constructed, by making identifications on
ordinary flat initial data, examples that again carry no $\cal D$-constant
spinor fields, but now definitely have nontrivial \mcal.
In such examples, the mass
function $M_N$ will be strictly positive. What is it? Is it true that,
for every slice in Minkowski space-time, \mcal\ is 2-dimensional?
For every Cauchy surface?

Must the infimum in the definition, (17), of the norm mass
function always be realized? That is, must there always exist,
for any $\alpha\in$ \mcal, representative $\lambda^A$ with
$M_N(\alpha)=\parallel\!\lambda\!\parallel^2$?
It is easy to show that the infima are always realized on any
initial-data set that is ``generic'' in the sense that there is
a point of it at which $\rho > |\rho^a|$. Furthermore, existence of an
infimum would follow in every case from Conjecture 7.
Nevertheless, realization of an infimum does
not follow in general from elementary
facts about operators on a Hilbert space: It is not hard to construct
an example, on a Hilbert space, of a positive-semi-definite Hermitian
quadratic form $\zeta$ and a translate $W$ of a closed subspace
such that the infimum of $\zeta$ on $W$ is not realized.

We have seen in Theorem 4 that, for any asymptotically flat
initial-data set, \mcal\ has the structure of a spinor space: It is
2-dimensional, with an adjoint operation $\dagger$ and an alternating
tensor $\epsilon$. So, we introduce the real, 4-dimensional vector
space $\cal V$ with its Lorentz metric $g$ and preferred unit timelike
vector $t$, on which the mass function $M_N$ becomes a real linear
function. How
much of all this can be carried over to more general initial-data
sets? We can in every case introduce $\cal V$ as the self-adjoint
elements of the tensor product of \mcal\ and its complex-conjugate
space. There results a real vector space (in infinite dimensions,
a Hilbertable one, reflecting that \mcal\ is Hilbertable), on which
the mass functions $M_N$ and $M_A$ in every case become real linear
functions. But what of the remaining structure on \mcal? We cannot
guarantee an adjoint operation as in Theorem 4: It is false
for a general initial-data set that every element of \mcal\
has representative $\lambda^A$ with $\lambda^{\dagger A}\in$ \hcal\
(e.g., that of the hyperboloid in Minkowski space-time). Howerer,
there are always such representatives when $p_{ab}$ has compact
support, and so, presumably, under suitable asymptotic conditions
on $p_{ab}$. Is there a simple theorem to this effect?
It seems to be more difficult to obtain an alternating
tensor as in Theorem 4: It is not even close to being true, for
a general initial-data set, that any two elements of \mcal\ have
representatives with inner product asymptotically constant.
We remark that an adjoint operation on \mcal, with no alternating
tensor, gives rise to a certain linear mapping on $\cal V$ (in the
asymptotically flat case, a reflection about the $t$-axis); and
that an alternating tensor on \mcal, with no adjoint operation, gives
rise to a metric on $\cal V$. Perhaps there is some other structure,
combining parts of $\dagger$ and $\epsilon$, that can always be defined.

What happens to the space \mcal\ and the mass function $M_N$ and
$M_A$ under time-evolution of an initial-data set? The simplest case
is that in which evolution takes place only within a compact
subset of $T$. Then, as we have seen in Sect. 3, neither the
space \mcal\ nor the asymptotic mass function $M_A$ changes.
The example of Eqn. (52) strongly suggests that, in
general, the norm mass function will change under this evolution.
What of the space $Z$ --- the vector space of solutions of Eqn. (46)?
It seems likely that the dimension of this space, at least, will not
change under evolution. Indeed, if $Z$ is 2-dimensional, then the data
is for a flat spacetime, and so, therefore, will be any evolution
of those data. If $Z$ is 1-dimensional, then the spacetime has null Weyl
tensor and matter a null fluid. This character is probably also
preserved under time-evolution. Thus, we suggest, the dimensionality of
$Z$ should be an evolution-invariant. Evolution within a compact
subset of $T$ can proceed eventually to singular behavior. An example
is that of a slice in the extended Schwarzschild spacetime, evolving to
reach ``$r=0$''. Is it true that \mcal\ and $M_A$ are preserved even
under {\it this} evolution?
That is, does the present framework ignore ``internally generated''
singular behavior? To prove that it does would require
good control over that singular behavior.
What happens for evolution not restricted to
compact sets? In the asymptotically flat case, we know that the
space \mcal\ and the mass functions $M_N$ and $M_A$ all
remain invariant. Is there any similar result
using conditions significantly weaker than asymptotic flatness?

\section{Acknowledgement}

This work was supported in part by National Science Foundation
Grant No. PHY--9220644 to the University of Chicago.

\bigskip

\noindent $^1$ R. Arnowitt,  S. Deser,  and C. W. Misner,
{\it Gravitation: An Introduction to Current Research}, ed Witten, L.
(New York: Wiley, 1962)

\noindent $^2$ R. Penrose,  Proc. R. Soc. Lond. {\bf A381}, 53--63 (1982).

\noindent $^3$ A. J. Dougan,  and L. J. Mason,  Phys. Rev. Lett.
{\bf 67}, 2119--22 (1991);
G. Bergqvist,  Class. Quan. Grav. {\bf 9}, 1917--22 (1992).

\noindent $^4$ R. Bartnik,   Phys. Rev. Lett. {\bf 62}, 2346--8 (1989).

\noindent $^5$ J. D. Brown, and J. W. York,  Phys. Rev. D
{\bf 47}, 1407--19 (1993).

\noindent $^6$ M. Ludvigsen, and J. A. G. Vickers, J. Phys. A
{\bf 16}, 1155-68 (1983).

\noindent $^7$ E. Witten,  Commun. Math. Phys. {\bf 80}, 381--402 (1981).

\noindent $^8$ We have set $4\pi G=1$.

\noindent $^9$ See Ref. 7 and A. Sen,
J. Math. Phys. {\bf 22}, 1781--6 (1981).

\noindent $^{10}$ One might think that there could be introduced
additional mass functions, intermediate between
$M_N$ and $M_A$,
as follows. Set
$$
{ M_\tau(\alpha)=\inf \Bigl (\parallel\!\lambda\!\parallel^2
+\tau \int |{\cal D}_A^{\ \ B}\lambda_B |^2 \Bigr ) }
$$
where the infimum is over all $\lambda^A$
in equivalence class $\alpha$,
and where $\tau \ge -2$
(which guarantees that the infimum exist).
Then $M_{-2}=M_A$ and
$M_0=M_N$. But this yields nothing new, for
we have, using the fact that the right side of (18) is independent
of representative, $M_\tau=-\tau/2 M_A +(1+\tau/2)M_N$.

\noindent $^{11}$ By this, we mean that each of
$M_A$ and $M_N$ satisfies
the polarization relation: For every
$\alpha ,\ \beta\in {\cal S}$,
$M(\alpha+\beta)+M(\alpha - \beta)=2[M(\alpha)+M(\beta)]$.
This is immediate for $M_A$.
For $M_N$, it follows from the following fact:
For any $\alpha\in {\cal S}$, and
$\lambda^A$ any representative, we have
$\parallel\lambda\parallel^2-M_N(\alpha) \le \epsilon^2$
if and only if $|<\!\lambda|\sigma\! >|
\le \epsilon \parallel\sigma\parallel$
for every $\sigma\in {\cal C}$.

\noindent $^{12}$ We note a few technical points regarding this theorem.
We actually need less than that $T=R^3$, namely that
$T$ outside a compact set be diffeomorphic with
$R^3$ outside a compact set.
And the Euclidean metric $\qz_{ab}$
need only be specified outside that compact set. Hypothesis
$i)$ actually follows from hypotheses $ii)$
and $iii)$, together
with the condition that $\qz_{ab}$
bound $q_{ab}$ below.
Indeed, it even follows from these that $q_{ab}-\qz_{ab}$
approaches zero asymptotically to order $r^{-1/5}$.
Hypotheses $i)$ and $ii)$
alone suffice to show that there
is some 2-dimensional subspace of \mcal\ satisfying conclusions
$ii)$ and $iii)$.
Hypotheses $iii)$ is then needed only to
show that this subspace exhausts \mcal. Were we assuming power-law
asymptotic behavior of $p_{ab}$
and $\Dz_aq_{bc}$ --- which we
emphatically are not --- then hypothesis
$iii)$ would follow already from $ii)$.
Hypothesis $iii)$ can in any case be replaced by
the weaker hypothesis that $r|p_{ab}|$
and $r|\Dz_a q_{bc}|$ are
bounded asymptotically by sufficiently small --- but
nonzero --- numbers. Conclusion $ii)$ actually holds for
every representative $\lambda^A$
of $\alpha\in$\mcal.
Furthermore, a weaker notion of ``approach'' can be
incorporated into conclusion $iii)$ so that it,
too, holds for all representatives.

\noindent $^{13}$
A tensor field on $T$ is said to approach zero asymptotically
provided that, given any $\epsilon >0$, there exists a compact
subset of $T$ outside which the
$q$-norm of that field is
everywhere less than $\epsilon$.

\noindent $^{14}$ See Ref. 7 and O. Reula,  J. Math. Phys.
{\bf 23}, 810-14 (1982).

\noindent $^{15}$
R. Bartnik, Comm. Pure and Appl. Math. {\bf 39}, 661--93 (1986).

\noindent $^{16}$
In fact, there is a simple analog of condition $i)$ in
Newtonian gravitation. The Newtonian mass density,
$\rho$, is a function on
Euclidean space $T$. Following case $i)$
above, let there exist a positive function
$\kappa$ on Euclidean space such
that the $B_{\kappa_0}$'s of Theorem 5 are all compact,
and such that $|D\kappa/\kappa|^2$
bounds $\rho$ below.
Then, we claim, the total Newtonian mass, $\int_T \rho dV$,
must be infinite. Indeed, were $\rho$ integrable, then so would
be $|D\log \kappa |^2$,
whence, by Theorem 2, $\log \kappa$
would approach a constant at infinity, violating compactness
of the $B_{\kappa_0}$'s.

\noindent $^{17}$ R. D. Adams,  {\it Sobolev Spaces} (Academic Press,
N. Y., 1975), theorems 4.28, 7.48 and 7.58.

\noindent $^{18}$
It can occur that, even though ${\lambda^A\in Z}$ is nonzero,
this $\lambda^A$ gives rise to
the zero point of ${\cal S}$. See, e.g.,
the earlier example
of flat data on $S^1\times S^1\times R$.

\noindent $^{19}$
There is in fact a simple Euclidean version of Conjecture 7.
Conjecture: Let $\vec v$ be a smooth complex vector field,
defined on the closed unit ball $B$ in Euclidean
$R^3$.
Then there exists a number $c>$0 with the following property:
Given any smooth complex function $f$ on $B$,
there is a function $f_0$ on $B$, there satisfying
$\vec \nabla f_0 = \vec v f_0$, such that
$$
\int_B{|\vec \nabla f - \vec v f|^2} \ge c \int_B{|f-f_0|^2}.
$$

\noindent $^{20}$
By this, we mean that, for any $\alpha=\alpha_1+\alpha_2$, with
$\alpha\in {\cal S},\ \alpha_1\in {\cal S}_1,\ \alpha_2\in {\cal S}_2$,
we have $M_A(\alpha)=M_{1A}(\alpha_1)+M_{2A}(\alpha_2)$.

\noindent $^{21}$ R. Penrose, Proc. R. Soc. Lond.
{\bf A284}, 159 (1965), R. Geroch,
in Asymptotic Structure of Space-time,
F. P. Esposito, L. Witten, Ed. (Plenum, N. Y. 1977).

\noindent $^{22}$ O. Reula and K. P. Tod, J. Math. Phys. {\bf 25},
1004--8 (1984).

\noindent $^{23}$ Note that the Lorentz group acts as data-preserving
diffeomorphisms on $T$, yielding an action of this
group as mass-preserving linear maps on \mcal.

\end{document}